\documentclass[a4paper]{jpconf}
\usepackage{graphicx}

\begin{document}
\global\long\def\gbar{\ensuremath{\bar{g}}}

\title{\rightline{\small\rm IIT-CAPP-16-1}\\[-0.1in]
Antimatter gravity with muonium$^*$}

\author{Daniel M Kaplan$^{1,a}$, Ephraim Fischbach$^2$, Klaus Kirch$^3$, Derrick C Mancini$^1$, James D Phillips$^4$, Thomas J Phillips$^1$, Robert D Reasenberg$^{4,b}$, Thomas J Roberts$^{1,c}$ and Jeff Terry$^1$}

\address{$^1$ Illinois Institute of Technology, Chicago, IL 60616,  USA}
\address{$^2$ Purdue University, West Lafayette, IN 47907, USA}
\address{$^3$ Paul Scherrer Institute, Villigen and ETH, Z\"urich, Switzerland}
\address{$^4$ Harvard-Smithsonian Center for Astrophysics, Cambridge, MA 02138, USA}
\address{$^a$ Presenter}
\address{$^b$ Current address, Center for Astrophysics and Space Sciences, University of California, San Diego}
\address{$^c$ Also at Muons, Inc.}

\ead{kaplan@iit.edu}

\begin{abstract}
The gravitational acceleration of antimatter, \gbar, has never been directly measured and could bear importantly on our understanding of gravity, the possible existence of a fifth force, and the nature and early history of the universe. Three avenues appear feasible  for such a measurement: antihydrogen, positronium, and muonium. The muonium measurement requires a novel monoenergetic, low-velocity, horizontal muonium beam directed at an atom interferometer. The precision three-grating interferometer can be produced in silicon nitride or ultrananocrystalline diamond using state-of-the-art nanofabrication. The required precision alignment and calibration at the picometer level also appear to be feasible. With 100\,nm grating pitch, a 10\% measurement of \gbar\ can be made using some months of surface-muon beam time, and a 1\% or better measurement with a correspondingly larger exposure. This could constitute the first gravitational measurement of leptonic matter, of 2nd-generation matter and, possibly, the first measurement of the gravitational acceleration of antimatter.
\end{abstract}

\section{Antimatter gravity: a brief introduction}
While {\em indirect} tests imply stringent limits on the gravitational acceleration of antimatter ($\gbar/g-1<10^{-7}$)~\cite{Alves} (\gbar\ being the gravitational acceleration of antimatter on earth), such limits are based on the varying amounts of virtual antimatter presumed to contribute to nuclear binding energies in various elements. The extent to which they  apply to muonium is thus far from obvious. (Virtual antimuons surely play a negligible role in the nucleus.) 

A {\em direct} test of the gravitational interaction of antimatter with matter seems desirable on quite general grounds~\cite{Nieto-Goldman}. 
Such a measurement could bear importantly on the formulation of a
quantum theory of gravity and on our understanding of the early history
and current configuration of the universe. It may be viewed as a test
of general relativity (GR), or as a search for new, as-yet-unseen,
forces, and is of great interest from both perspectives.

\vfill\leftline{\footnotesize ~~$^*$Presented at the Third Workshop on Antimatter Gravity (WAG2015), London, England, 4--7 August 2015.}

GR predicts no difference between the gravitational behavior of antimatter
and that of matter. This follows from the geometric nature of GR and similarly from the equivalence principle, which
implies that the gravitational force on an object is independent of
its composition. While well established experimentally, GR is fundamentally
incompatible with quantum mechanics, and development of a quantum
alternative has been a longstanding quest. To date,  the available
experimental evidence on which to base such a  theory  derives only
from observations of matter--matter and matter--light interactions. Matter--antimatter
measurements could play a key role in this quest. Indeed, the most
general candidate theories include the possibility that the force
between matter and antimatter will be different from that of matter
on matter (for example, suppressed scalar and
vector terms may cancel in matter--matter interactions, but add in
matter--antimatter ones)~\cite{Nieto-Goldman}.

Although most physicists expect the equivalence principle to hold
for antimatter as well as for matter, theories in which this symmetry
is maximally violated (i.e., in which antimatter ``falls up'') are
attracting increasing interest~\cite{DK1}-\cite{DK15}
as potentially providing alternative explanations for six great mysteries
of physics and cosmology: why is the cosmic-ray background radiation
so isothermal, why is the universe so flat, what happened to the antimatter,
where is the dark matter, what is the dark energy, and why does the universe appear younger than the oldest stars? As an example
of the range of theoretical possibilities, a recent paper \cite{Tasson}
examining a possible standard-model extension (SME) discusses a model
that evades restrictions from existing measurements while allowing
differing gravitational response for matter and antimatter.  By pointing
out that ``the SME coefficients {[}...{]} can differ between sectors,
so investigations of higher-generation matter-gravity couplings are
of independent interest,'' the authors emphasize the importance of
2nd-generation gravitational measurements. 
Given the short lifetimes of second- and third-generation particles,
muons may provide the only experimentally accessible direct measurements
of gravity beyond the first generation.  

While a number of experimental approaches have been attempted or proposed, none has yet achieved significance: the only published direct limit to date, on antihydrogen, is $-65 <  \gbar/g < 110$~\cite{ALPHA}. Despite being some orders of magnitude from the desired realm of sensitivity, given the technical challenges it is a good start, and the collaboration is pursuing ideas for improvement~\cite{Lyman-alpha-laser}.
Besides antihydrogen, only two other experimental avenues appear practicable: measurements on positronium~\cite{Hogan} and on muonium~\cite{Kirch} (M), a $\mu^+e^-$ hydrogenic atom. 

\section{Measuring muonium gravity}

We are developing a precision three-grating  atom-beam interferometer for the measurement of \gbar\ using a slow beam of muonium under development at  Switzerland's Paul Scherrer
Institute (PSI). Such a measurement will constitute a unique test of the gravitational interaction of leptonic and second-generation matter with the gravitational field of the earth. To
measure the free fall of muonium, a low-velocity horizontal beam of
M atoms will be aimed at an atom interferometer. The interferometer
can measure the  atoms' gravitational deflection  to a fraction
of a nanometer. This approach can determine \gbar\ to a precision better than 10\%
of $g$ in some months of beam time  at PSI. Additional beam time, intensity, or efficiency could then permit a \gbar\ measurement
to 1\% or better. 
The enabling technologies for this measurement are  an ultra-precise
atom-beam interferometer and a slow, monochromatic beam of muonium.   We have begun the needed interferometer R\&D
using teams of undergraduates at IIT. The muonium beam R\&D is ongoing at PSI, with test runs scheduled $\approx$\,annually~\cite{Eggenberger}.

\section{Principle of the measurement}
\begin{figure}
\centerline{\includegraphics[width=.75\linewidth]{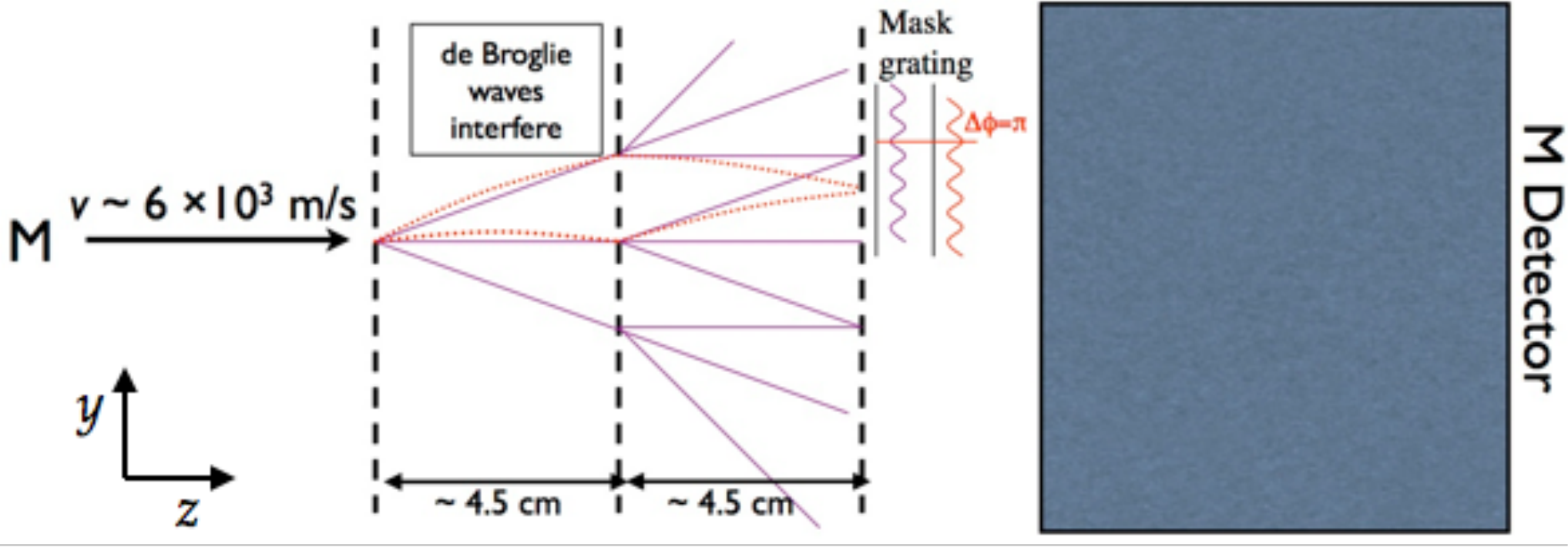}}
\caption{Principle of muonium interferometer, shown in elevation-view cross section (3-lifetime grating spacing and phase difference $\Delta\phi= \pi$ shown for illustrative purposes). De Broglie waves due to each atom diffract at first and second gratings producing interference pattern, sampled by vertically scanning the third grating.}\label{fig:interf}
\end{figure}

The muonium measurement requires a precision three-grating interferometer (Fig.~\ref{fig:interf}), coupled with a low-background muonium detection system. The first two gratings create an interference pattern, which is analyzed by displacing the third grating vertically, using piezoelectric actuators, and detecting and recording the rate of muonium passing through it vs.\ its displacement.
The dimensions of the interferometer will be selected so as to minimize the combined statistical
and systematic uncertainty. In one muon lifetime ($\tau = 2.2\,\mu$s) the standard-gravity muonium deflection is $\frac{1}{2}g\tau^2 = 24$\,pm, yielding an interferometric phase shift 
$\phi = 2\pi g\tau^2/d = 0.003$ if (as proposed by Kirch~\cite{Kirch}) gratings with $d = 100$\,nm pitch, separated by one muon lifetime, were used. Measuring $\phi$ to $< 10$\% would then require grating fabrication fidelity, and interferometer stabilization and alignment, at the picometer level\,---\,a substantial technological challenge. Larger grating separations give a larger phase shift but, due to muon decay, reduced event statistics. The  separation that minimizes the statistical uncertainty is two lifetimes, so the separation that will minimize the combined statistical and systematic uncertainty will be somewhat larger than two lifetimes. For example, if a grating separation of three lifetimes is used, the same number of muonium atoms incident on the interferometer gives comparable statistical significance as with a separation of one muon lifetime, but with nine-times larger deflection and phase shift, requiring ``only'' tens of pm tolerances. Achieving such tolerances is expected to be within the current state of the art~\cite{Thapa-etal}, although some R\&D will be necessary in order to adapt the established techniques to our geometry.
The RMS statistical precision of the measurement is estimated as~\cite{Kirch,Oberthaler}
\begin{equation}
\delta g=\frac{1}{C\sqrt{N}}\frac{d}{2\pi}\frac{1}{t^2}\,,
\end{equation}
where $C = 0.1$ is the estimated fringe contrast, $N$ the number of events detected, and $t$ the muonium traversal time through the interferometer. At the anticipated rate of $10^5$ M atoms/s incident on the interferometer, and taking into account decays and estimated inefficiencies, the statistical measurement precision is about $0.3g$ per $\sqrt{N_d}$, $N_d$ being the exposure time in days. All other things being equal, a finer grating pitch would appear to be helpful. The chosen 100\,nm pitch reflects a compromise between sensitivity and systematic error due to grating variations over the $\sim$\,cm$^2$ grating area. 

\section{Apparatus design}
The interferometer consists of three equally spaced transmission gratings, all having the same pitch (a Mach--Zehnder configuration~\cite{Chang-etal}), each of which must be precisely aligned to the others. As shown in Fig.~\ref{fig:interf}, in such an interferometer, the 0th and $\pm$1st diffraction orders from the first grating are diffracted again by the second grating so that they recombine and interfere at the location of the third grating. With 50\%-open gratings, even orders are suppressed, and the three diffraction orders shown contain most of the transmitted intensity in approximately equal amounts. The resulting interference pattern is sinusoidal with period equal to the grating pitch. Since each atom's de Broglie wave interferes with itself, and the interference patterns from all atoms are in phase with each other, this interferometer configuration accommodates an extended, incoherent source, and the alignment of the beam axis with the interferometerÕs axis is not critical~\cite{Chang-etal}.

Although based on similar principles as previous atom-beam interferometers, 
the precision and source size for the proposed experiment demand improvements beyond those previously built. We note that, assuming (for the sake of illustration) \gbar\ $ = g$ and 3-lifetime grating separation, the phase difference corresponds to about 0.4\,nm of gravitational deflection. 
The relative transverse alignment of the interferometer gratings will therefore need to be controlled with an accuracy of about 0.04\,nm for a 10\% measurement and 4\,pm for a 1\% measurement.
We believe we can meet these specifications in both accuracy and precision using the semiconductor-laser tracking frequency gauges (TFGs) described below. It will also be vital to minimize and monitor long-term drift in this phase measurement since the statistics needed will be accumulated over a period of days or longer. While piezeoelectric nanopositioning actuators are capable of the required position resolution, their absolute position readout is of more limited (typically about 1\,nm) accuracy. We therefore rely on precise measurement and feedback in order to meet these specifications.

Most mechanical drift is caused by thermal expansion. We can minimize such fluctuations by mounting the gratings on a stable, rigid structure: an optical bench machined from a single block of silicon. Silicon's coefficient of thermal expansion is extremely small at low temperatures (4\,pm/m$\cdot$K at 2\,K, 100\,pm/m$\cdot$K at 6\,K) 
and (as discussed below) the interferometer will be designed to be compatible with use in a cryostat at superfluid-helium temperature. The expansion coefficient is also zero near 125\,K, a more convenient temperature at which to work while evaluating the stability of prototypes. Channel-cut silicon crystals are an established technology for use in cryogenic monochromators for synchrotron radiation facilities~\cite{Zhang-etal}. They have been specifically engineered and employed for operation near 125\,K to minimize thermal expansion effects when thermally loaded by exposure to synchrotron radiation. We will utilize this established technology for the optical bench. 

Since it will not be possible to achieve perfect alignment of the 100-nm-period gratings upon assembly, two of the gratings will need to be aligned to each other using piezoelectric actuators, and the third will need to be scanned in order to make the phase measurement. For use with atomic beams, the interferometer must be capable of operating in vacuum. Vacuum-compatible piezo-driven actuators are commercially available and work well at cryogenic temperatures~\cite{Martinet-etal}.

\begin{figure}
\includegraphics[width=.5\linewidth]{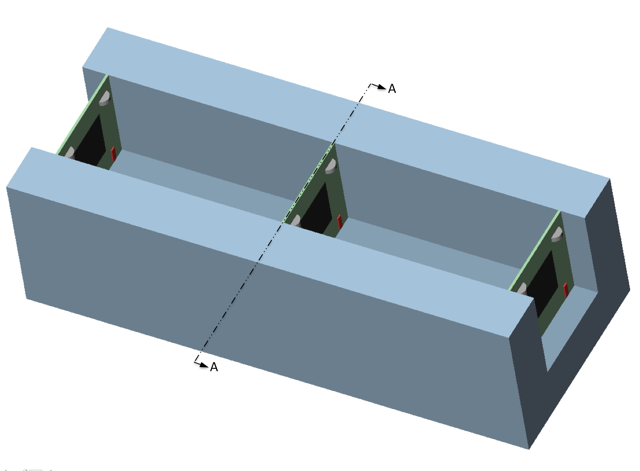}\includegraphics[width=.5\linewidth]{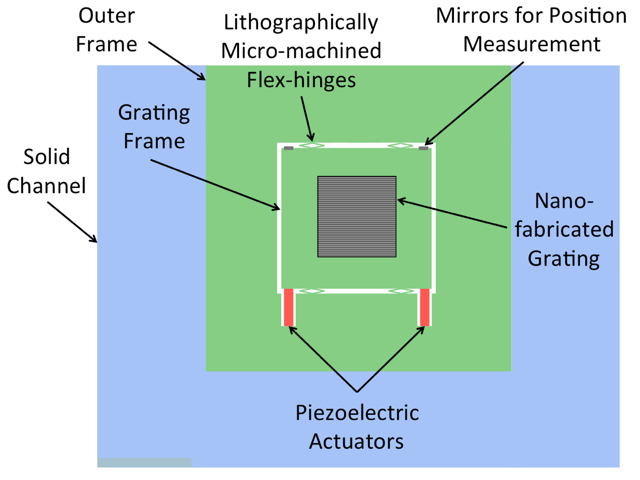}
\caption{(left) CAD drawing of the muonium interferometer; (right) Section A-A. In blue-gray is the support structure for the gratings, a solid channel machined out of a single silicon block. Each grating is mounted in a silicon frame connected to an outer frame by flex-hinges; a pair of piezoelectric actuators permits small rotations to align the gratings precisely in parallel, as well as scanning of the third grating. Each grating frame has mirrors (or possibly corner-cube retroflectors) on its top corners that form part of the optical TFGs used to measure their position.}\label{fig:3D}
\end{figure}

A 3D CAD drawing of the interferometer is shown in Fig.~\ref{fig:3D}. The first grating is fixed to the support structure and establishes the coordinate system to which the other gratings must be aligned. The other gratings will be mounted on frames that can be moved vertically using piezoelectric actuators in order to align their slits parallel to those of the first grating. These actuators will also be used to scan the relative phase of the gratings, with the interference pattern inferred from the resulting sinusoidal modulation of the muonium counting rate. (We note that rotations of the gratings about coordinate axes other than the beam axis result in $\cos{\theta}$ effects, so that extreme alignment precision about those axes is not required.) 

We require the vertical noise motion of the interferometer to be small compared to a grating period over the $\approx$\,13\,$\mu$s flight time of the muonium through the interferometer. RMS noise with an amplitude of 10\% of the grating period would reduce the interference contrast by only about 1\%, which is acceptable. (Note that this estimate also sets the scale of acceptable imprecision in grating
fabrication.) Typical seismic noise peaks at 1\,$\mu$m\,(RMS)/$\sqrt{\rm Hz}$ at 0.1\,Hz, falling exponentially to 1\,nm/$\sqrt{\rm Hz}$ at 1\,Hz, beyond which it remains flat to 10\,Hz. It then falls as $1/f^2$ at higher frequencies, so seismic noise is well below the level of concern. On the other hand, some laboratory equipment such as turbopumps vibrate at problematic frequencies, so we will need acoustic isolation. The maximum p-p grating motion during passage of a muonium atom results from vibration at 39 kHz. Fortunately, vibration amplitudes tend to be small at this frequency, and isolation is very effective.
Suspending the optical bench on wires, sized to be stressed to 30\% of yield tension, gives a vertical bounce resonance at $\approx10$\,Hz. This forms an acoustic low pass filter ($1/f^2$), with $\approx10^2$ rejection at 100\,Hz and $10^4$ at 1\,kHz, adequate to suppress vertical motion for a $10^{-2}$ \gbar\ measurement. Horizontal motions are more suppressed due to the lower frequency of the pendulum resonance, $\approx1$\,Hz. (Yet further isolation could be implemented but is most likely not necessary.)

Muonium production in thin stopping targets gives typical muonium velocities of $\sim$\,6\,mm/$\mu$s. Thus, the three-lifetime grating separation discussed above corresponds to $\sim$\,4.5\,cm, as shown in Fig.~\ref{fig:interf}. This small extent of the interferometer eases stabilization, as does the short muonium traversal time, since vibration periods much longer than microseconds do not affect the measurement. Nevertheless, as already noted, alignment and stabilization at the tens-of-picometer level will be challenging.

It is important to have low background and to count only muonium atoms that have passed through the interferometer. This can be accomplished by employing a coincidence technique. The decaying muons will emit positrons, which (due to the high $Q$-value) emerge preferentially at large angles to the muonium beam direction. These can be counted in a segmented scintillation counter or a compact crystal calorimeter surrounding the beam path downstream of the interferometer; if needed, positron tracking is also possible~\cite{Regenfus}. The remaining, no-longer bound, electrons can be electrostatically accelerated towards a charged-particle detector such as a microchannel plate~\cite{Rosen-etal}. The coincidence of these two signals can be used to suppress background due to cosmics and beam muon decays in the vicinity of the interferometer. 

In order to establish the interferometer alignment for an undeflected beam to the needed sub-nm precision, soft x-rays of wavelength comparable to that of the muonium atoms (0.6\,nm) will be used. The measurement will thus determine the phase difference between the x-ray beam interference pattern and that due to the muonium beam. We can use a position-sensitive MCP for the muonium detector's electron counter; this will also detect soft x-rays. The position measurement has a number of advantages: it will
allow us to see Moir\'{e} interference fringes to help in aligning the gratings, and it will allow us to match the muonium and x-ray illumination patterns to minimize any systematic bias between their respective interference patterns.

A novel approach has been proposed for formation of the muonium beam and is under development by our collaborators at PSI: stopping of surface muons in a $\sim$\,$\mu$m-thick film of superfluid helium (SFHe)~\cite{Taqqu}. The efficacy of SFHe for muonium production is already experimentally established~\cite{Abela-etal}. Due to the (mass-dependent) chemical potential of hydrogenic atoms within SFHe, muonium formed within the film should be ejected approximately perpendicular to the SFHe surface at a velocity of 6.3\,mm/$\mu$s with a quite low (0.2\%) energy spread. (The chemical potential calculations have been experimentally verified for deuterium~\cite{Reynolds-etal}.) The beam thus formed could be reflected~\cite{Luppov-etal} into the horizontal by a thin SFHe film coating an angled section of the cryostat wall oriented at 45$^\circ$. This approach 
would require the interferometer and detector to be integrated, and operated, within a cryostat at SFHe temperature. The monoenergetic beam would ease measurement of the free-fall deflection, since all M atoms would spend the same amount of time traversing the interferometer, and their interference patterns would thus be in phase with each other. (With a less monochromatic beam, a chopper or other velocity-selection or -measurement  technique would be needed.)

\section{R\&D program}
We will explore three options to measure the grating phase position. The simplest is to monitor the voltages applied to the piezoelectric actuators.
We will measure the relative phase of the gratings using x-ray interference, which will also provide a measure of the stability and repeatability of the piezoelectric actuators. We will also develop the application of TFGs to interferometer calibration and verification.

Semiconductor-laser tracking frequency gauges, developed by J. D. Phillips and his colleagues at the Harvard-Smithsonian Center for Astrophysics (CfA), are capable of sub-picometer length measurements~\cite{Thapa-etal}. Because calibration is provided by the x-ray phase, the TFGs need measure only changes in relative grating position, and be stable only over the time required for x-ray calibration. For this purpose, optical interferometers, formed by a mirror or cornercube endpoint mounted on a grating and another on a fixed support, will be used; these interferometers
can be Fabry--Perot cavities or non-resonant. Each optical interferometer will have a laser locked to one of its nulls. When an endpoint moves, its interferometer changes length and the laser locked to it changes frequency. Relative position shifts are determined with extreme accuracy by measuring beat frequencies among pairs of locked lasers. We will start with two donated TFGs: they will allow us to characterize the stability of the muonium interferometer, to calibrate the piezoelectric drive, and to monitor grating translation to extremely high precision. 

For absolute alignment, we propose a unique approach that exploits the interferometer's ability to use soft x-rays, produced at a synchrotron radiation source (or, in the final experiment, using a bench-top source), of similar wavelength as anticipated for the muonium beams (0.6\,nm). These measurements, which can be interleaved with the muonium measurement for real-time alignment tracking, determine the absolute, zero-deflection, phase since the traversal time, and hence the gravitational phase shift, of x-rays through the interferometer is negligibly small. Changes in the absolute phase monitor the system's drift. For measurement of the muonium gravitational phase shift, the stability requirement is that these drifts be slow on the time scale needed to measure the absolute phase with x-rays. In the final experiment, the TFG measures short-term variations and the x-rays provide long-term calibration. (The x-ray flux must be limited, in part due to the thermal load on the gratings.) During development, x-ray and TFG measurements will be compared to learn whether there are any significant structural distortions not measured by a small number of TFGs. We do not expect any, but if there are, remedies include adding TFGs to monitor the additional degrees of freedom.

We have already obtained approval from the Center for Nanoscale Materials (CNM) at Argonne National Laboratory for grating fabrication, and have started fabrication of prototype gratings, in collaboration with CNM personnel. The structures would ideally be 100-nm-period grids of horizontal slits. For mechanical stability, however, buttresses between lines of the grating are required, producing a grid of mini-slits that approximates an ideal grating. The grid structures will be made in a thin film of silicon nitride or ultrananocrystalline diamond (UNCD), possibly coated with higher-$Z$ metal, as membrane windows on silicon wafers. In addition to fabricating the grid structures, optimizing process and choice of material, metrology and inspection will also be carried out at the CNM. The expected outcome is a set of grids that can be used for initial testing with x-ray and atomic beams, and verification of stability compared to modeling calculations performed using finite element analysis (FEA). The grid structures of 50-nm-wide slots will be laid out using L-Edit software and 
patterned using $e$-beam lithography. The pattern will be transferred into a thin film of silicon nitride or UNCD on silicon wafers using reactive ion etching (RIE). Back windows will be etched into the supporting silicon wafers using KOH wet etching and optical lithography patterning. 
Test exposures for the grid structures using the $e$-beam lithography tool have been made, demonstrating feasibility of using the tool for the required dimensions and size. We plan to incorporate  flexural hinges into the silicon frame around the grating by using micromachining (optical lithography and RIE or wet etching). 

We are developing the needed muonium beam collaboratively with PSI. While, at present, interest in muonium gravity is strongest at PSI, alternative venues could also be considered. Fermilab does not now possess a surface-muon beam, but options for such beams are discussed in the Project X report~\cite{ProjectX} and were presented at the CSS2013 summer-study meeting~\cite{Plunkett}; a low-intensity test surface-muon beam at Fermilab for R\&D may also be possible~\cite{Plunkett-priv}. The Project X intensity estimate, $\sim$\,10$^{10}$\,muons/s~\cite{ProjectX}, is comparable to what has been discussed for a future facility at PSI. 
Of course, the future accelerator path for Fermilab is somewhat uncertain at present. While their average surface-muon beam rates are lower, TRIUMF, RAL, and J-PARC represent additional venues at which muonium R\&D and experimentation could be carried out.

\section{Conclusions}
An experiment to measure the gravitational acceleration of muonium atoms is currently in an  R\&D stage. Such a measurement would be complementary to other antimatter gravity tests in progress or proposed. While significant technological challenges exist, they appear far from insuperable.

\section*{Acknowledgments}
We have benefited from discussions with the late Andy Sessler and from the support of the IPRO program~\cite{IPRO} at Illinois Institute of Technology. We thank the Smithsonian Astrophysical Observatory (part of the Harvard-Smithsonian Center for Astrophysics) for a signiÞcant donation to IIT of equipment, including two TFG laser gauges. The development of a suitable muon and muonium beam is supported by the Swiss National Science Foundation, grant \#200020\_146902.


\section*{References}
\begin{thebibliography}{99}
\bibitem{Alves}
Alves D S M, Jankowiak M and Saraswat P 2009 Experimental constraints on the free fall acceleration of antimatter {\it Preprint} arXiv:0907.4110 [hep-ph]
\bibitem{Nieto-Goldman}
Nieto M M and Goldman T 1991
The Arguments Against ``Antigravity'' and the Gravitational Acceleration of Antimatter {\it Phys. Rep.} {\bf 205} 221
\bibitem{DK1}
Blanchet L 2007 Gravitational polarization and the phenomenology of MOND {\it Class. Quant. Grav.} {\bf 24} 3529
\bibitem{DK2}
Blanchet L and Le Tiec A 2008 Model of dark matter and dark energy based on gravitational polarization {\it Phys. Rev. D} {\bf 78} 024031
\bibitem{DK3}
Burinskii A 2008 The Dirac-Kerr-Newman electron 
{\it Grav. Cosmol.} {\bf 14} 109
\bibitem{DK4}
Blanchet L and Le Tiec A 2009 Dipolar dark matter and dark energy {\it Phys. Rev. D} {\bf 80} 023524
\bibitem{DK5}
Hajdukovic D S 2011 Is dark matter an illusion created by the gravitational polarization of the quantum vacuum? {\it Astrophys. Space Sci.} {\bf 334} 215
\bibitem{DK6}
Villata M 2011 CPT symmetry and antimatter gravity in general relativity {\it EPL} {\bf 94}(2) 20001
\bibitem{DK7}
Benoit-L\'evy A and Chardin  G 2012 Introducing the Dirac-Milne universe {\it Astron. Astrophys.} {\bf 537} A78
\bibitem{DK8}
Hajdukovic D S, 2012 Quantum vacuum and dark matter {\it Astrophys. Space Sci.} {\bf 337} 9
\bibitem{DK9}
Hajdukovic D S, 2012 Quantum vacuum and virtual gravitational dipoles: the solution to the dark energy problem? {\it Astrophys. Space Sci.} {\bf 339} 1
\bibitem{DK10}
Villata M 2013 On the nature of dark energy: the lattice Universe {\it Astrophys. Space Sci.} {\bf 345} 1
\bibitem{DK11}
Benoit-L\'evy A and Chardin  G 2014 The Dirac-Milne cosmology {\it Int. J. Mod. Phys.: Conf. Series} {\bf 30} 1460272
\bibitem{DK12}
Hajdukovic D S 2014 Virtual gravitational dipoles: The key for the understanding of the universe? {\it Phys. Dark Univ.} {\bf 3} 34
\bibitem{DK13}
Dallal S A, Azzam W J and de Fez M 2015 Gravitational signature of matter-antimatter interaction, {\it J. Mod. Phys.} {\bf 6} 201.
\bibitem{DK14}
Villata M 2015 The matter-antimatter interpretation of kerr spacetime {\it Ann. Physik} {\bf 527} 507
\bibitem{DK15}
Phillips T J 2016 Antimatter may matter {\it Nature} {\bf 529} 294
\bibitem{Tasson}
Kostelecky V A  and Tasson J D  2011 Matter-gravity couplings and Lorentz violation {\it Phys. Rev. D} {\bf 83} 016013
\bibitem{ALPHA}
Amole  C {\it et al} (ALPHA collaboration) 2013 Description and first application of a new technique to measure the gravitational mass of antihydrogen {\it Nature Commun.} {\bf 4} 1785
\bibitem{Lyman-alpha-laser}
 ALPHA collaboration plans for greater sensitivity are discussed in Hamilton P {\it et al} 2014 {\it Phys. Rev. Lett.} {\bf 112} 121102
\bibitem{Hogan}
Hogan S 2015 Rydberg positronium for tests of antimatter gravity {\it  This workshop}
\bibitem{Kirch}
Kirch K 2007 Testing Gravity with Muonium {\it Preprint} physics/0702143 (2007)
\bibitem{Eggenberger}
Eggenberger A 2015 Towards a novel high-brightness muon beamline for next generation precision experiments {\it This workshop}
\bibitem{Thapa-etal}
Thapa R, Phillips J D, Rocco E and Reasenberg  R D  2011 Subpicometer length measurement using semiconductor laser tracking frequency gauge {\it  Opt. Lett.} {\bf 36} 3759
\bibitem{Oberthaler}
Oberthaler M K 2002 Anti-matter wave interferometry with positronium {\it Nucl. Instrum. Methods B} {\bf 192} 129
\bibitem{Chang-etal}
Chang B J, Alferness R and  Leith E N 1975 Space-invariant achromatic grating interferometers: theory {\it Appl. Opt.} {\bf 14} 1592
\bibitem{Zhang-etal}
Zhang L {\it et al} 2003 The performance of a cryogenically cooled monochromator for an in-vacuum undulator beamline {\it J. Synchr. Rad.} {\bf 10} 313
\bibitem{Martinet-etal}
Martinet G {\it et al} 2006 Low temperature properties of piezoelectric actuators used in SRF cavities cold tuning systems {\it Proc. EPAC 2006 (Edinburgh)} p 390
\bibitem{Regenfus}
Regenfus C 2003 A cryogenic silicon micro-strip and pure-CsI detector for detection of antihydrogen annihilations {\it Nucl. Instrum. Methods A} {\bf 501} 65 
\bibitem{Rosen-etal}
Ros\'en S {\it et al} 2007 Operating a triple stack microchannel plate-phosphor assembly for single particle counting in the 12--300 K temperature range {\it Rev. Sci. Inst.} {\bf 78} 113301 
\bibitem{Taqqu}
Taqqu D 2011 Ultraslow muonium for a muon beam of ultra high quality {\it Phys. Procedia} {\bf 17} 216
\bibitem{Abela-etal}
Abela E {\it et al} 1993 Muonium in liquid helium isotopes {\it JETP Lett.} {\bf 57} 157
\bibitem{Reynolds-etal}
Reynolds M W, Hayden M E and Hardy W N 1991 Hyperfine resonance of atomic deuterium at 1 K {\it J. Low Temp. Phys.} {\bf 84} 87
\bibitem{Luppov-etal}
Luppov V G {\it et al} 1993 Focusing a beam of ultracold spin-polarized hydrogen atoms with a helium-film-coated quasiparabolic mirror {\it Phys. Rev. Lett.} {\bf 71} 2405
\bibitem{ProjectX}
Asner D {\it et al} 2013 {\it Project X broader impacts} (Fermilab Report)\\
{\tt http://projectx-docdb.fnal.gov/cgi-bin/ShowDocument?docid=1200}
\bibitem{Plunkett}
Plunkett R K 2013 Surface muon beam at Project X {\it Presented at CSS2013 Workshop (Minneapolis-St. Paul)}\\
{\tt https://indico.fnal.gov/contributionDisplay.py/pdf?contribId=465\&sessionId=67\&confId=6890}
\bibitem{Plunkett-priv}
Plunkett R K {\it Private communication}
\bibitem{IPRO}
IIT Interprofessional Projects website: {\tt http://ipro.iit.edu/}
\end{thebibliography}
\end{document}